\documentclass[12pt,a4paper,final]{iopart}

\usepackage{bm}
\usepackage{color}
\usepackage{footnote}
\usepackage{graphicx}
\usepackage[breaklinks=true,colorlinks=true,linkcolor=blue,urlcolor=blue,citecol
or=blue]{hyperref}

\begin{document}

\title{Moving Beyond Chi-Squared in Nuclei and Neutron Stars}

\author{A.~W. Steiner}
\address{Institute for Nuclear Theory, University of
Washington, Seattle, WA 98195}
\eads{\mailto{steiner3@uw.edu}}

\begin{abstract}
There are several assumptions made in a standard $\chi^2$ analysis of
data, including the frequent assumption that the likelihood function
is well approximated by a multivariate Gaussian distribution. This
article briefly reviews the standard approach and describes how
Bayesian inference can be used to go beyond the assumption that the
likelihood is Gaussian. Two separate types of analysis relevant to
nuclear physics are used as test cases. The first is the determination
of the equation of state of dense matter from neutron star mass and
radius data. The second is the use of theoretical nuclear mass models
to fit currently available data and predict the value of masses which
have not yet been measured. For the problem of predicting nuclear
masses, it is demonstrated that approximating the likelihood function
with a Gaussian can produce biased predictions of unmeasured masses.
Finally, the lessons learned from these fitting problems are used to
propose a method for improving constraints on the nuclear symmetry
energy.
\end{abstract}

\vspace*{-0.2in}

\vspace*{0.2in}


\vspace*{0.2in}
\noindent {\small {\it J. Phys. G.: Nucl. Part. Phys.} {\rm in press}}

\noindent {\small{Preprint: INT-PUB-14-018}}

\section{Introduction}

Interpreting data and understanding the accompanying theoretical
models often requires a statistical analysis. Frequently, this
analysis takes the form of an minimization of a function, $\chi^2$,
which quantifies the extent to which a model can reproduce a set of
experimental data. Uncertainties and correlations are determined
by exploring the properties of the likelihood function near the
best-fit parameter set. This process contains several assumptions,
one of which is that the likelihood function is nearly Gaussian.

Neutron stars, to a good approximation, all lie on one, universal,
mass-radius curve. It is of broad interest to determine this universal
relation, in part because of the connection to the cold and dense
region of the QCD phase diagram~\cite{Lattimer01}. Statistical
analyses were not applied to neutron stars until
recently~\cite{Ozel09,Steiner10}, in part because of the paucity of
mass and radius data. The data is sufficiently sparse that the dense
matter equation of state is currently underconstrained. This has
motivated the use of Bayesian inference which provides a clearer path
forward for underconstrained systems. Systematic uncertainties also
play a dominant role, and Bayesian inference is useful for
characterizing these uncertainties~\cite{Kass95}.

The statistical analysis of nuclear mass data, on the other hand, was
performed already in the 1920s, just before the discovery of the
neutron and the creation of the Bethe-Weisacker mass
formula~\cite{Lunney03}. Historically, works have typically chosen to
minimize the root-mean-square deviation either of the mass excess, the
binding energy, or the energy per baryon~(see the discussion in
Ref.~\cite{Toivanen08}) with respect to parameters in a Hamiltonian
(e.g. the Skyrme model~\cite{Skyrme59} or a covariant mean field
model~\cite{Lalazissis97}), or a phenomenological nuclear mass
formula~\cite{vWeizsacker35,Bethe36,Duflo95,Dieperink09}. More recent
works have considered other data in the fits, such as charge radii
(e.g. Ref.~\cite{Lalazissis97}) and dipole
polarizabilities~\cite{Piekarewicz12}.

In section \ref{sec:chi2} the basic problem of fitting nuclear data is
introduced, along with a discussion of how Bayesian inference can play
a role. Section \ref{sec:ns} describes the analysis of neutron star
data, and shows how the Gaussian approximation fails to accurately
describe correlations and biases predictions of the radius of a 1.4
$M_{\odot}$ neutron star. Section \ref{sec:pred} describes predictions
of unmeasured nuclear masses and the nuclear symmetry energy, and
section \ref{sec:comp} highlights the importance of selecting the
proper data set.

\section{Chi-squared, nuclear masses, and charge radii}
\label{sec:chi2}

The prototypical problem to be discussed is the fitting of a model,
i.e. an energy density functional, to nuclear masses and charge radii.
A review the basic formalism highlights the various assumptions which
are often made. Using a set of $N_d$ data points, $D_i$, and a model with
$N_p$ parameters, $p_j$, one defines a function, $\chi^2$, by
\begin{equation}
\chi^2 = \sum_i^{N_d} \left[\frac{D_i-P_i(\boldsymbol{p})}
{\sigma_{\mathrm{exp},i}}\right]^2 \, ,
\end{equation}
where $P_i$ is the model prediction for data point with index $i$ and
$\sigma_{\mathrm{exp},i}$ is the experimental uncertainty of the
$i$-th data point. Minimizing $\chi^2$ over the $N_p$ dimensional
parameter space gives the best fit to the data. Alternatively, one
maximizes a likelihood function
\begin{equation}
{\cal L} = \exp\left(-\chi^2/2\right) \, ,
\label{eq:ldef}
\end{equation}
which is equal to a product of Gaussians for each data point. The
formulation into a likelihood function makes it clear that an implicit
assumption of an independent and Gaussian distribution for each data
point has already been made.\footnote{The assumption of independence
  is violated in nuclear mass data; the construction of the atomic
  mass evaluation is done in such a way that neighboring nuclear
  masses are non-trivially correlated~\cite{Audi86,Audi03}. Future
  atomic mass evaluations will attempt to assess these
  correlations~\cite{Kondev14}. In any case, these correlations may
  not be important here because the experimental uncertainties are
  much smaller than the systematic uncertainties. For examples of
  fitting correlated data in hadron mass spectroscopy, see
  Ref.~\cite{Michael94}.} If this assumption is correct and the
deviations of the data from the model predictions follow a Gaussian
distribution, then the value of $\chi^2$ follows the chi-squared
distribution and a ``good fit'' is one where $\chi^2 \approx N_d -
N_p$. Note that, in an underconstrained system, $N_d - N_p < 0$ and
this test is not useful.

If the model is a linear function of the parameters, then the model
serves as a linear transformation of the multivariate Gaussian
determined by the data and therefore the likelihood function is also
Gaussian. For nonlinear models, the likelihood function is
approximately Gaussian in the limit $N_d \rightarrow \infty$ because
of the central limit theorem. If the likelihood function is
approximately Gaussian, then it can be parameterized by a best fit vector
$\boldsymbol{\mu}$ and a covariance matrix, $\boldsymbol{\Sigma}$:
\begin{equation}
{\cal L}(\boldsymbol{p}) \approx A 
\exp \left[ -\frac{1}{2} \left(\boldsymbol{p}-\boldsymbol{\mu}\right)^T
\boldsymbol{\Sigma}^{-1}
\left(\boldsymbol{p}-\boldsymbol{\mu}\right)\right] \, ,
\label{eq:likenorm}
\end{equation}
with a normalization factor $A$ fixed so that Eq.~\ref{eq:ldef}
holds.\footnote{Note that the likelihood function is not renormalized
  so that its integral is unity. The likelihood function, unlike the
  prior and posterior distributions below, is not a probability
  density function. The normalization is important when comparing how
  two models fit to the data. See also the discussion of Bayes factors
  below.} The covariance matrix is often determined by computing the
second derivative of ${\cal L}$ around the maximum. In terms of the
original $\chi^2$ function,
\begin{equation}
\Sigma_{ij} = \frac{1}{2} \frac{\partial}{\partial p_i}
\frac{\partial}{\partial p_j}  \left( \chi^2 
\right)_{\boldsymbol{p}=\boldsymbol{\mu}} \, .
\label{eq:trad_covar}
\end{equation}
The correlation matrix $\boldsymbol{{\cal C}}$ can be obtained from
$\boldsymbol{\Sigma}$ by ${\cal C}_{ij} =
\Sigma_{ij}/\sqrt{\Sigma_{ii} \Sigma_{jj}}$. Modern fitting codes
often obtain the covariance matrix from the Jacobian of $\chi^2$. The
correlation matrix, obtained in this way, is a correct representation
of the correlations between model parameters {\em only if the
  likelihood function is approximately given by the form in
  Eq.~\ref{eq:likenorm}}. It has become commonplace to report
correlation matrices without verifying that this approximation is
actually valid. The implications of this approximation for
predicting nuclear masses is discussed in section \ref{sec:pred}
below. 

Alternatively, the covariance matrix can also be obtained directly
from a Markov chain Monte Carlo simulation of the likelihood function.
Each entry is just the correlation coefficient
\begin{equation}
\Sigma_{ij} = \sum_{i=1}^{N_c} \frac{1}{N_c} \left(p_i - \bar{p}_i\right)
\left(p_j - \bar{p}_j\right) \, ,
\label{eq:full_covar}
\end{equation}
where ${\bar{p}_i}$ is the average value of the $i$-th parameter
and $N_c$ is the number of entries in the Markov chain. This
latter method is useful even when the likelihood function
is not approximately Gaussian. 

In the context of nuclear masses, there is evidence that the
likelihood functions implied by typical models are often not Gaussian.
Explicit calculation shows that the likelihood function from the
simple Bethe-Weisacker mass formula is multi-modal because of the
pairing contribution. Skyrme models~\cite{Skyrme59}, energy density
functionals which describe nuclear masses and charge radii, are linear
in parameters $t_0, \ldots, t_3$, but not linear in the parameter,
$\alpha$, which characterizes density-dependent two-nucleon force.
This non-linearity could mean that likelihood is not fully Gaussian,
though this has not been fully studied. Some evidence for non-Gaussian
behavior is present in the posterior probability distributions for the
Skyrme-like parameters in Fig. 1 of Ref.~\cite{Schunck14a} which are
fit to a large set of data comprising mostly of nuclear masses and
charge radii. Covariant mean-field models may also not generate
Gaussian likelihoods because of the non-linearity (which is required
to reproduce saturation) in the mean-field equation for the
scalar-isoscalar meson (e.g. Ref.~\cite{RocaMaza14}).

If the likelihood function was a multivariate Gaussian, one would
expect the minimization of $\chi^2$ to be numerically trivial (so long
as the likelihood was evaluated with sufficient numerical accuracy).
For example, the Broyden-Fletcher-Goldfarb-Shanno method is guaranteed
to converge if the objective function has a quadratic expansion near
the extremum. The result that the fit of the Skyrme-like parameter
$C_{1}^{\rho \nabla J}$ depends slightly on the initial guess in
Ref.~\cite{Wild14} (in a fit similar to that in Ref.~\cite{Schunck14a}
described earlier) means that, in contrast, this minimization is not
trivial. This is not surprising as the posterior distribution of this
same parameter in Fig. 1 in Ref.~\cite{Schunck14a} appears
multi-modal.

All currently available energy density functionals which describe
heavier nuclei have systematic uncertainties which are much larger
than the $\sigma_{\mathrm{exp},i}$. Thus, it is common to redefine
$\chi^2$ as
\begin{equation}
\chi^2 = \sum_i^{N_d} 
\frac{\left(D_i-P_i\right)^2}
{\sigma_{\mathrm{exp},i}^2+\sigma_{\mathrm{sys},i}^2} \, ,
\end{equation}
where $\sigma_{\mathrm{sys},i}$ is the systematic uncertainty for each
data point. Often $\sigma_{\mathrm{exp},i}$ is small enough that it
can be ignored. There are now $N_p+N_d$ parameters: the $N_p$ model
parameters and a parameter $\sigma_{\mathrm{sys},i}$ for each data
point. Equivalently, one can replace each $\sigma_{\mathrm{sys},i}^2$
with a weight, $w_i$, to be fixed by some recipe. There are several
ways to proceed. It is possible to perform a Bayesian analysis with
all $N_p+N_d$ parameters, but this is computationally
difficult.\footnote{I have found empirically that a Bayesian analysis
  in this form requires prior distributions for
  $\sigma_{\mathrm{sys},i}$ that fall off as $\exp(-\sigma^2)$ to to
  ensure the trivial solution with $\sigma \rightarrow \infty$ does
  not dominate the results. More work in this direction is in
  progress.} The alternative is to reduce the number of parameters. A
frequent choice is to employ a fixed uncertainty for each type of data
point, e.g. $\sigma_{\mathrm{mass}}$ for nuclear masses and
$\sigma_{\mathrm{radius}}$ for charge radii. A similar choice is
manifest in the literature through the presence of fitting protocols:
to choose not to fit data point $j$ is equivalent to assuming that the
systematic uncertainty for the model to predict the $j$-th data point,
$\sigma_j$, is so large that the corresponding term in $\chi^2$ is
negligable. Early fits (see review in e.g. Ref.~\cite{Lunney03}) used
only doubly-magic nuclei, on the basis that open-shell systems
contained correlations which were unlikely to be correctly computed by
Hartree-Fock calculations. Ref.~\cite{Dobaczewski14} discusses this
issue and suggests fixing the denominators by separately normalizing
$\chi^2$ for each type of data (their Eq. 4). A slightly different
procedure is suggested below. Finally, computing several systematic
uncertainties using iteratively-reweighted least squares may be
useful. Studies of how relative variations in the $\chi^2$ weights
might affect the interpretation of the data, in the context of
low-energy nuclear data, are in their infancy.

An implicit assumption in this discussion is the assumption
that the systematic uncertainties in the denominators are themselves
uncorrelated. If this assumption fails, then one must reformulate the
$\chi^2$ function to take into account the uncertainties. This problem
has been found particularly relevant for fitting parton distribution
functions (see in particular Appendix B of Ref.~\cite{Stump01}). If
correlations among parameters {\em} and correlations among the
systematic uncertainties are not important, and if the $\chi^2$
function is of the form given above, one can estimate parameter
uncertainties by varying the parameters, one at a time, until
$\chi^2(\boldsymbol{p}) = N_d - N_p + 1$. However, it seems likely
that the systematic uncertainties of masses and radii are indeed
correlated. Models with no three-nucleon forces often predict
saturation at a higher density and with a larger binding energy than
found in laboratory nuclei~\cite{Coester70}, and these models
naturally lead both to smaller masses and smaller radii.

In the context of fits to nuclear masses and charge radii, the model
parameters are almost always correlated. If the likelihood function is
nearly Gaussian, then one can determine the parameter uncertainty
including these correlations by marginalizing over the multivariate
Gaussian (which can be done using the covariance matrix and does not
require any integration).

\section{Bayesian analysis for meutron star masses and radii}
\label{sec:ns}

For neutron stars, the problem is how to determine the mass-radius
$M-R$ curve from a small sample of poorly constraining observations.
(A review is available in Ref.~\cite{Lattimer14a}, only some of which
is particularly relevant to this work and reproduced here.) The
statistical analysis is complicated for two reasons. The first is
that the currently available data has large statistical (and
large systematic) uncertainties. There are effectively more
parameters than constraints. Second, the mass-radius curve need not be
a function in the mathematical sense, it may fail either or both of
the horizontal or vertical line tests. (Given the central energy
density, $\varepsilon_c$, the relations $M(\varepsilon_c)$ and
$R(\varepsilon_c)$ are both well-defined functions.) An example of how
this happens in is Ref.~\cite{Alford13}, where a phase transition
induces a new branch in the $M-R$ curve.

Bayesian analysis allows one to solve both of these issues, at the
expense of introducing an unknown function, a prior probability
distribution function. Also, the two-dimensional nature of the neutron
star problem requires that $N_d$ parameters for the neutron star
masses, $M_i$, are also added. Some researchers have been slow to
employ Bayesian inference because of the ambiguity in determining
prior distributions, but Ref.~\cite{Gelman12} has shown that the prior
is not a separate entity but a part of the model which can be checked
and verified by comparing to data.

In the case of nuclear mass fitting described above, the likelihood
function is a product of $N_d$ Gaussian probability distributions for
each data point. The neutron star case is handled similarly (c.f.
Eq.~31 in Ref.~\cite{Steiner10}): the likelihood function is a product
over a data set for each neutron star. Each data set is of the form
${\cal D}(M,R)$, a two-dimensional probability distribution for the
mass and radius of a neutron star. These probability distributions
have complicated shapes, and the likelihood is not Gaussian. As
discussed above, the traditional correlation matrix thus does not
contain all of the information on correlations present in the problem.

\begin{figure}
\includegraphics[width=6.0in]{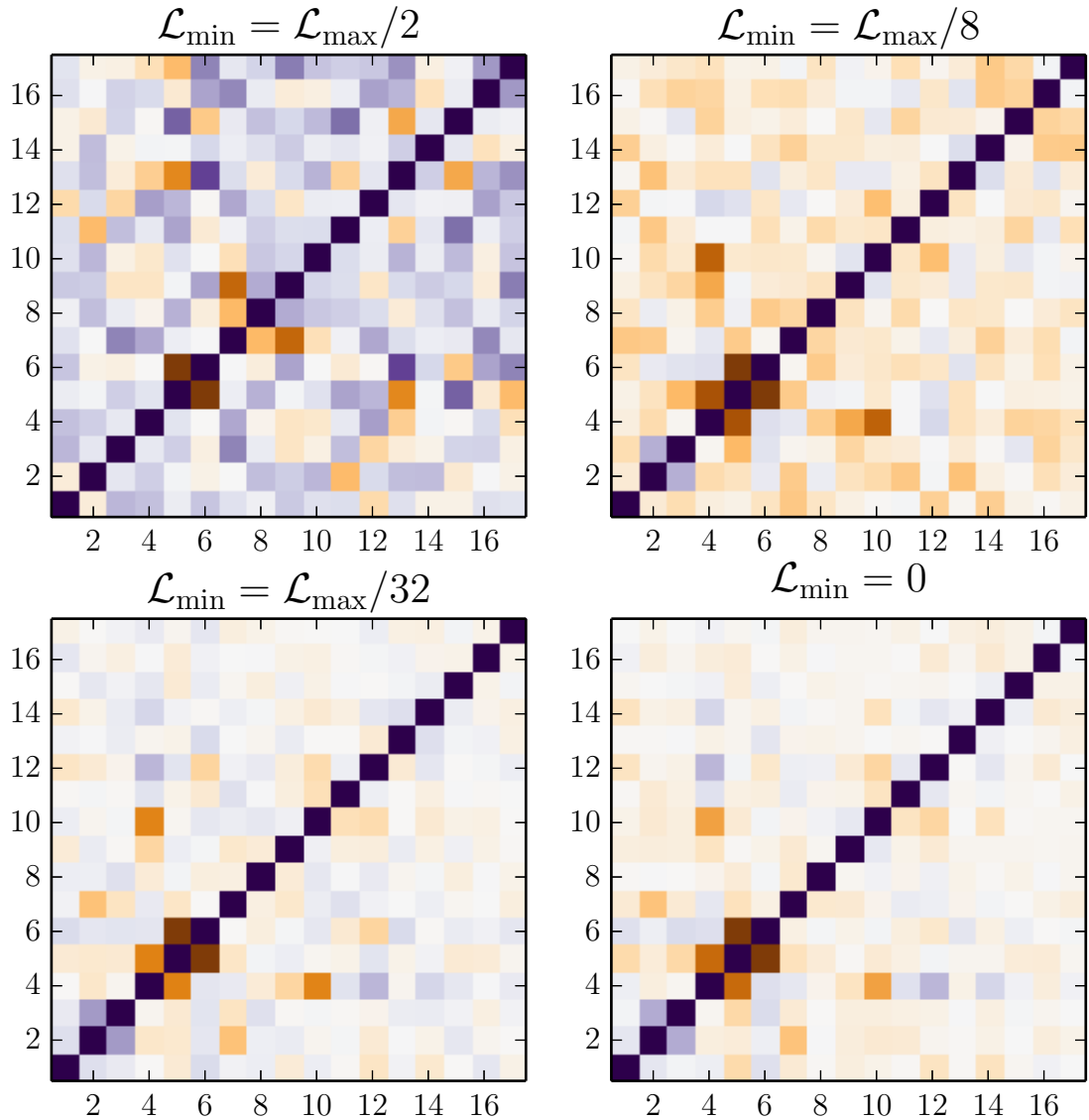}
\caption{The evolution of the correlation matrix as more Monte Carlo
  points near the maximum likelihood (best fit) are added. The x and
  y-axes both enumerate the parameter number as given in Table 1. The
  upper-left shows the correlation matrix in a close neighborhood
  around the best fit. This is the result as obtained assuming the
  likelihood is described by a covariance matrix from
  Eq.~\ref{eq:trad_covar}. The lower-right panel shows the full
  covariance with all the Monte Carlo samples (using
  Eq.~\ref{eq:full_covar}). The upper-right and lower-left panels show
  the evolution between these two. Strong correlations are in purple,
  yellow, strong anti-correlations are in orange, and lack of
  correlation is white. The true correlations in the fit are
  represented by the lower-right panel even though the upper-left
  panel is representative of the method most often used to compute
  these correlations.}
\label{fig:corr_mats}
\end{figure}

As a demonstration, the correlation matrix is represented in
Fig.~\ref{fig:corr_mats}, for the analysis of neutron star data with
the HLPS + Model C parameterization from Ref.~\cite{Steiner14} based
on Ref.~\cite{Hebeler13} and Model C from Ref.~\cite{Steiner13}. There
are 17 model parameters, 7 from the EOS and 10 neutron star masses.
The 7 EOS parameters are given in Table 1. The results below do not
depend sensitively on the values of the saturation density or the
energy density grid which is used, so long as the grid covers the full
range of energy densities probed in neutron star cores. The four
panels each represent the correlation matrices inferred from different
subsets of the Markov chain Monte Carlo simulation of the likelihood
function. The upper-left panel was computing using only points where
the likelihood was within a factor of two of the maximum value at the
best fit (this is the result which corresponds to the Gaussian
approximation as discussed in section~\ref{sec:chi2}). The upper-right
panel used points where the likelihood panel was a factor of 8 within
the maximum. The lower-right panel used all the Monte Carlo points and
thus does not impose any approximation on the likelihood function. The
strong correlation between parameters 2 and 3 is related to the
correlation between S and L as obtained in Ref.~\cite{Hebeler13}. The
strong anti-correlation between pressure parameters is also expected.
The pressure cannot be too small over a large range in density or the
maximum mass will like below the observed value. The pressure cannot
also be too large over a large range in density or the radii will be
too large to reproduce the data. Qualitatively, the pressure tends to
be either small at moderate densities and large at high densities, or
vice versa. Finally, it is clear that many of the correlations in the
Gaussian approximation (upper-left panel) are not representative of
the full data set (lower-right panel).

\begin{figure}
\includegraphics[width=3.2in]{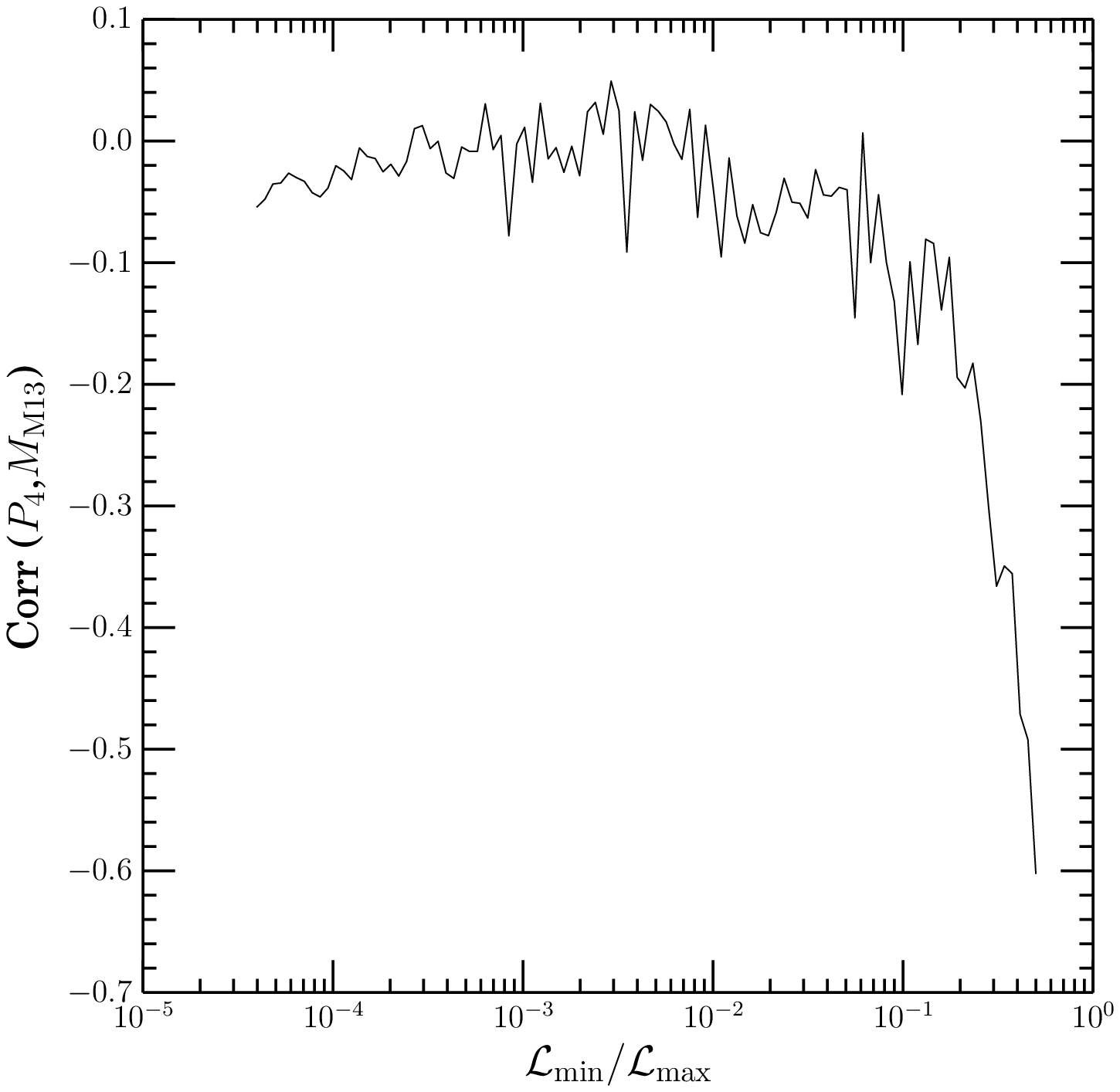}
\includegraphics[width=3.2in]{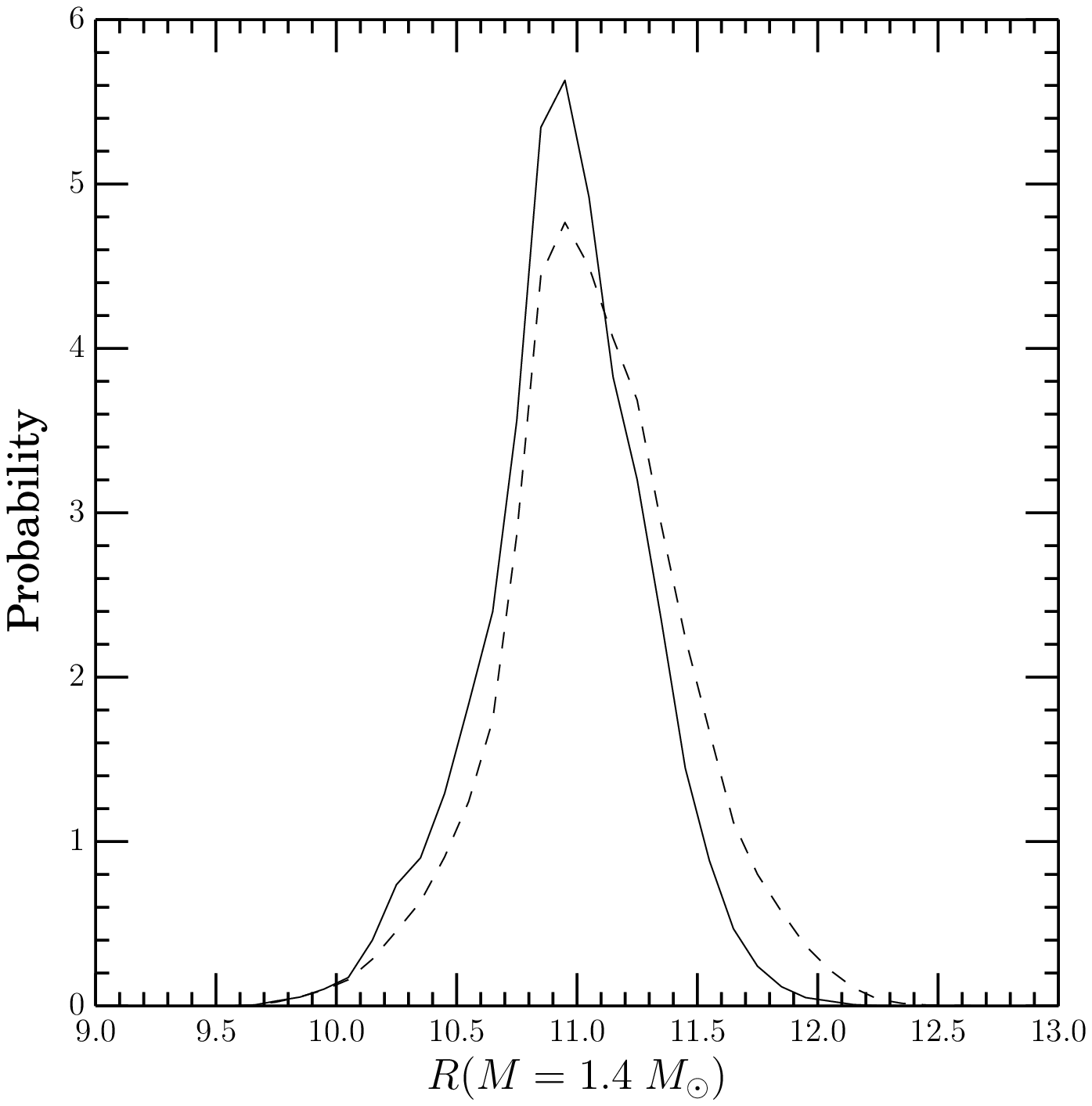}
\caption{Left panel: the evolution of one correlation coefficient away
  from the best fit. The pressure at high densities and the mass of
  one neutron star appear correlated (on the right side of the plot),
  but in reality are uncorrelated (on the left side of the plot).
  Right panel: the posterior probability distributions for the radius
  of a 1.4 $M_{\odot}$ neutron star, assuming the full likelihood
  function (solid line) or a Gaussian approximation (dashed line). }
\label{fig:corr_plots}
\end{figure}

This result is further demonstrated in the left panel of
Fig.~\ref{fig:corr_plots}, where the correlation coefficient between
the high density pressure parameter, $P_4$ and the mass of the neutron
star in the M13 globular cluster is shown as a function of the range
of likelihoods considered, ${\cal L}_{\mathrm{min}}/{\cal
  L}_{\mathrm{max}}$. The two parameters appear anti-correlated near
the best fit. In reality, they are not correlated. This lack of
correlation is expected, as the neutron star in M13 is typically lower
in mass, and thus less sensitive to the pressure at higher densities.
The right panel of Fig.~\ref{fig:corr_plots} shows that the Gaussian
approximation to the likelihood also gives a different range for the
radius of a 1.4 $M_{\odot}$ neutron star implying slightly larger
radii.

\begin{table}
\begin{tabular}{ll}
Parameter index & Description \\
\hline
1 & Compressibility \\
2 & Energy per baryon of neutron matter \\
3 & Derivative of energy per baryon of neutron matter \\
4 & $P_1 \equiv P(\varepsilon=2~\mathrm{fm}^{-4})-
P(n_B=0.16~\mathrm{fm}^{-3})$ \\
5 & $P_2 \equiv P(\varepsilon=3~\mathrm{fm}^{-4})-
P(\varepsilon=2~\mathrm{fm}^{-4})$ \\
6 & $P_3 \equiv P(\varepsilon=5~\mathrm{fm}^{-4})-
P(\varepsilon=3~\mathrm{fm}^{-4})$ \\
7 & $P_4 \equiv P(\varepsilon=7~\mathrm{fm}^{-4})-
P(\varepsilon=5~\mathrm{fm}^{-4})$ \\
8-17 & Individual neutron star masses \\
\hline
\end{tabular}
\caption{The list of parameters for neutron star EOS parameterization
described in the text and corresponding to the correlation matrices
displayed in Fig.~\ref{fig:corr_mats}.}
\end{table}

In Bayesian inference, marginalization is often used to determine
model parameters. The posterior probability distribution for the
$i$-th parameter is determined from
\begin{equation}
P(p_i) \propto \int dp_1 dp_2 \ldots d_{p_{i-1}} d_{p_{i+1}} \ldots
dp_{N_p} dM_1 \ldots dM_{N_d} {\cal L}(\boldsymbol{p},\boldsymbol{M}) 
{\cal P}(\boldsymbol{p},\boldsymbol{M})
\end{equation}
where ${\cal L}$ is the likelihood and ${\cal P}$ is the prior
distribution. This integral is typically evaluated using a Monte Carlo
method. One may directly parameterize the $M-R$ curve and compare it
to the data or parameterize the EOS of dense matter, $P(\varepsilon)$
and then use the TOV equations to compute the $M-R$ curve at each
point in the Monte Carlo integration. Each parameterization (given a
fixed data set) is a different model. A particular advantage of this
formalism, as applied in
Ref.~\cite{Steiner10,Steiner12,Steiner13,Lattimer14a,Lattimer14b,Steiner14},
is that one need not assume any correlations between the properties of
matter near the saturation density and matter in the inner core of the
neutron star. Matter in the core may be best described, for example,
by quark degrees of freedom and have little similarity with neutrons
and protons at the saturation density.

Models can be compared with each other by computing the Bayes factor.
The Bayes factor of Model A with respect to Model B is the ratio,
$B_{AB} \equiv E_A/E_B$ where $E$ is the evidence:
\begin{equation}
E = \int d\boldsymbol{p}~d\boldsymbol{M}~{\cal L} 
(\boldsymbol{p},\boldsymbol{M}) 
{\cal P}(\boldsymbol{p},\boldsymbol{M})
\end{equation}
Note that the two models need not have the same number of parameters.
The Bayes factor is equivalent to betting odds: $B_{AB}=10$ implies
that model A is ten times more likely than model B. 

Several different models can be compared with Bayes factors, a
strategy employed in Ref.~\cite{Lattimer14a}. This work used Bayes
factors and an analytical model of the neutron star atmosphere to show
that different values of the ``hydrogen column density'' were favored.
The hydrogen column density is a parameter which characterizes the
extent to which X-rays are absorbed between the neutron star and the
detector. Ref.~\cite{Lattimer14a} also found a model where some
neutron stars have Helium atmospheres was favored. Recent work in
Ref.~\cite{Heinke14}, including more observational data, has confirmed
a lower hydrogen column density for the neutron star in the $\omega$
Cen globular cluster and a Helium atmosphere for the neutron star in
NGC 6397, as partially predicted in Ref.~\cite{Lattimer14a}.

\section{Predictions, the nuclear symmetry energy, and the 
neutron skin thickness of lead}
\label{sec:pred}

Given any function ${\cal F}(\boldsymbol{p})$ of the model
parameters\footnote{Because this section is more general, the
explicit reference to the neutron star masses, $\boldsymbol{M}$, is 
removed.}, one can use the best fit to compute the predicted value,
${\cal F}(\boldsymbol{p}=\boldsymbol{\mu})$. To compute the
uncertainty in the prediction, one can compute a probability
distribution
\begin{equation}
P(F) = 
\int d\boldsymbol{p} {\cal L} (\boldsymbol{p}) 
\delta\left[{\cal F}(\boldsymbol{p})-F\right] {\cal P}(\boldsymbol{p})
\label{eq:pred}
\end{equation}
This integral is easier if ${\cal L}$ can be accurately described by a
multivariate Gaussian, because one can directly sample the likelihood
function using the Cholesky decomposition of the covariance matrix.
When the Gaussian approximation is inadequate, as in the neutron star
problem discussed above, and the likelihood function must be
indirectly sampled, e.g. through Markov chain Monte Carlo. 
In order to make the Monte Carlo more efficient,
the Gaussian approximation can be used as a proposal distribution
in a Metropolis-Hastings step.

\begin{figure}
\begin{center}
\includegraphics[width=3.4in]{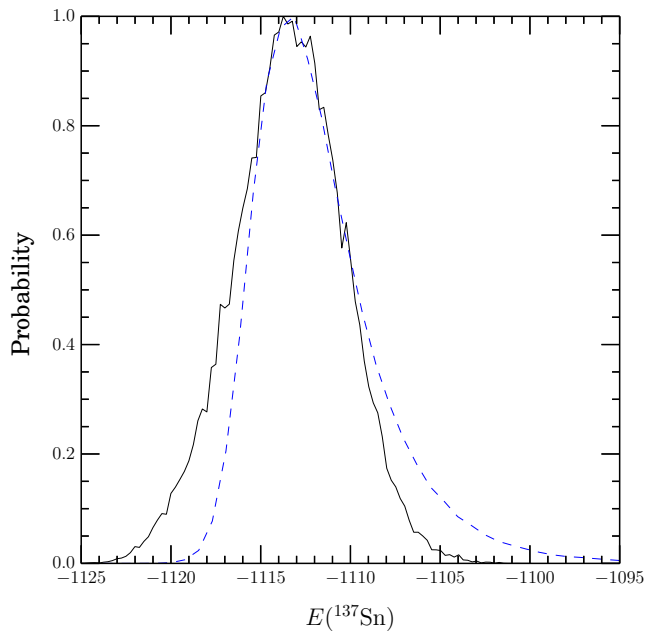}
\end{center}
\caption{Predicted probability distribution for the mass of
$^{137}\mathrm{Sn}$ given the nuclear mass model of
  Ref.~\cite{Dieperink09}. The black solid line gives the full result
from Eq.~\ref{eq:pred} using a Monte Carlo simulation without
any approximation and the blue dashed line gives the result 
using a Gaussian approximation for ${\cal L}$. The latter 
underestimates the predicted binding energy.}
\label{newfig}
\end{figure}

Fig.~\ref{newfig} demonstrates how the Gaussian approximation of the
likelihood function discussed in section~\ref{sec:chi2} affects
predictions for nuclear masses. The nuclear mass model of
Ref.~\cite{Dieperink09} is fit to the nuclear masses from the UNEDF
collaboration~\cite{Kortelainen12}. Eq.~\ref{eq:pred} is used to
predict the binding energy of $^{137}\mathrm{Sn}$. Uniform prior
distributions are used to allow a straightforward comparison to what
would be obtained in the frequentist method. Two results are
presented, with and without the Gaussian approximation of the
likelihood function. It is clear that the Gaussian approximation
systematically predicts less binding. 

A quantity of central interest in low-energy nuclear physics is the
nuclear symmetry energy, the energy cost to create an isospin
asymmetry. Given the energy per baryon of neutron matter as a function
of the baryon density, $E_{\mathrm{neut}}(n_B)$, and the energy per
baryon of nuclear matter $E_{\mathrm{nuc}}(n_B)$, the symmetry energy
can be defined as the difference $S(n_B) \equiv
E_{\mathrm{neut}}(n_B)-E_{\mathrm{nuc}}(n_B)$. Of particular interest
is the value of the symmetry energy at the nuclear saturation density,
$S \equiv S(n_B=n_0)$, and its derivative, $L \equiv 3 n_0
S^{\prime}(n_B=n_0)$.

Arguably, two of the best ways to obtain constraints on the nuclear
symmetry energy are from nuclear data and neutron star observations
described in sections~\ref{sec:chi2} and \ref{sec:ns} above. Nuclear
masses offer a strong constraint on a linear combination of $S$ and
$L$, but do not determine the two quantities
separately~\cite{Farine78}. Neutron star radius measurements and
measurements of the neutron skin thickness of lead both strongly
constrain $L$ and are weakly correlated with $S$~\cite{Horowitz01}.

In part motivated by the connection between the neutron skin thickness
in lead, $\delta R$, and the parameter $L$, several recent experiments
have measured the skin thickness. The PREX experiment, which uses
parity violating electron scattering, found $\delta
R=0.33^{+0.16}_{-0.18}$ fm~\cite{Abrahamyan12}. This weak probe of
nuclear structure should have smaller systematic uncertainties than
those from strongly interacting probes. A more precise (but more
model-dependent) result of $\delta R=0.15^{+0.04}_{-0.06}$ fm was
recently obtained from photoproduction of pions at the MAMI electron
beam facility in Ref.~\cite{Tarbert14}. Using probability
distributions determined by neutron star radius measurements, several
authors have determined the parameter
$L$~\cite{Steiner10,Hebeler10,Steiner12,Steiner13}.
Ref.~\cite{Steiner13} predicted the neutron skin thickness would be
less than 0.20 fm, consistent with both the recent PREX and MAMI
measurements. In this work, fits employed four seperate models (named
A, B, C, and D) and different interpretations of the neutron star data
in order to attempt to assess the systematic uncertainty (see Fig. 4
of Ref.~\cite{Steiner13}). It remains to be seen if the prediction of
the neutron skin thickness of lead from neutron star observations in
Ref.~\cite{Steiner13} will prove true with more accurate experiments,
such as PREX II.

In general, uncontrolled systematic uncertainties are particularly
pernicious for predicted values. {\em There is no method which
  guarantees that reality lies between any computed confidence
  interval which is obtained from the data}. In the context of energy
density functionals, there is a long history of using several nearly
equivalent models of the same data in order to estimate the systematic
uncertainties, as done in the neutron star analysis above, as earlier
done in Refs.~\cite{Furnstahl02,Stone03,Steiner05} and as discussed in
Ref.~\cite{Dobaczewski14}.\footnote{These studies can be viewed as an
  inexpensive and rough way of performing a hierarchical Bayesian
  analysis. In particular, they are only representative of the
  systematic uncertainty so long as they faithfully represent the
  space of reasonable models.} Studies of inter-model dependence are
important, even when one is not focused on predicting a particular
observable, because they help diagnose cases where a model may fit the
data accurately with an incorrect physical mechanism.

\section{Completeness versus accuracy}
\label{sec:comp}

The purpose of making a model is two-fold: the first is to provide
insight regarding the physical mechanisms which underlie the observed
data, and second is to make predictions. These two purposes are
sometimes at odds: the ability of a model to make predictions must
often strike a balance between attempting to describe the most
complete set of currently available data, and describing a small set
of data with high accuracy. This qualitative picture is behind the
practice (in Bayesian inference) of using different prior
distributions for parameter estimation than those used for model
comparison.

Much of the recent work surrounding the description of nuclear
structure observables has focused on generating an energy density
functional. The Kohn-Sham theorem suggests that, if the correct energy
density functional was found, one could accurately describe all nuclei
at the level of single reference Hartree-Fock-Bogoliubov (see e.g.
Ref.~\cite{Kortelainen12}). In many works, the purpose is mainly
that of completness: one describes as large of a data set as is
possible in order to obtain the best energy density functional. Part
of the promise of chiral effective theory is that we might not have to
sacrifice completeness for accuracy (see review in
Ref.~\cite{Epelbaum09}), and progress is being made in this direction
in lighter nuclei. On the other hand, very accurate descriptions of
neutron-rich calcium isotopes have been obtained with coupled cluster
methods~\cite{Hagen12,Wienholtz13} with interactions which might otherwise have
difficulty describing nuclear matter at the saturation
density~\cite{Hagen13}.

\begin{figure}
\includegraphics[width=3.2in]{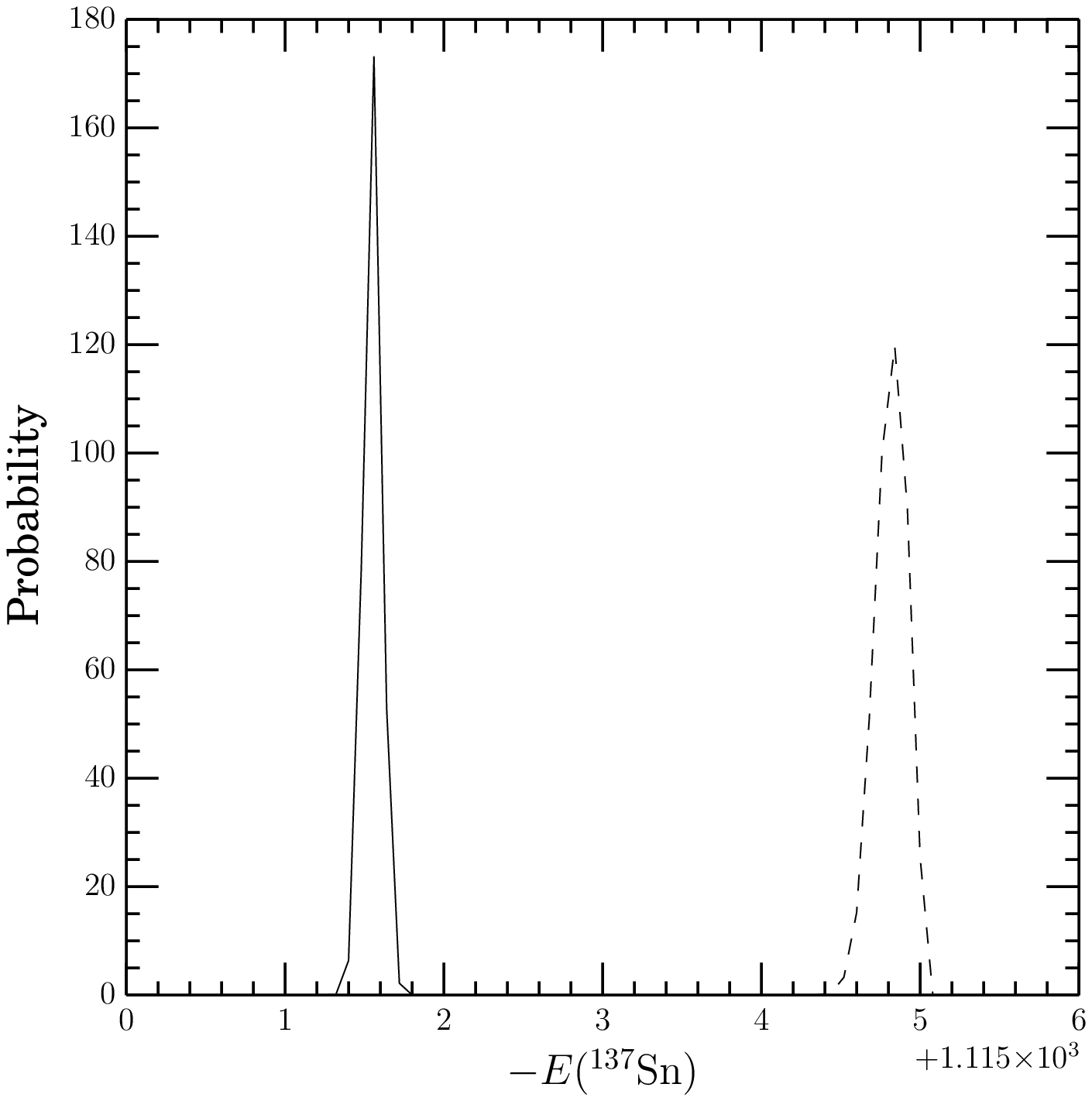}
\includegraphics[width=3.2in]{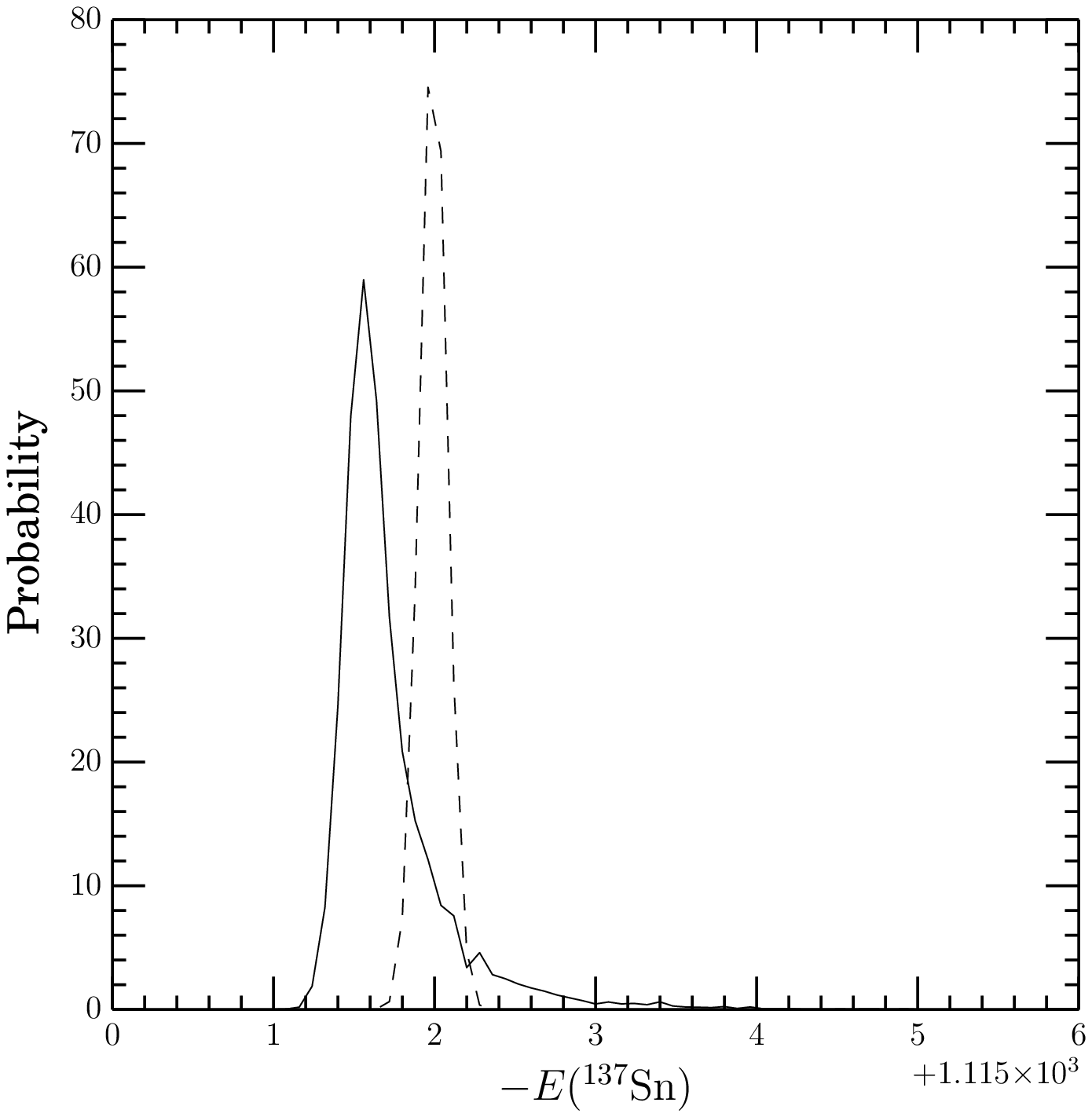}
\caption{A demonstration of the competition between completeness and
  accuracy. Solid lines use the model from Ref.~\cite{Dieperink09} and
  dashed lines use the mass model from Ref.~\cite{Steiner12b}. Left
  panel: the predicted binding energy of $^{137}$Sn when fitting all
  measured nuclear masses. Right panel: the binding energy of
  $^{137}$Sn when fitting only Sn, In, and Sb isotopes.}
\label{fig:data_comp}
\end{figure}

In the context of nuclear masses, this tension between completeness
and accuracy is demonstrated in Fig.~\ref{fig:data_comp}, where two
mass models from Refs.~\cite{Dieperink09,Steiner12b} are used to
predict the mass of $^{137}$Sn. In the left panel, the full
experimental mass data set from Ref.~\cite{Audi12,Wang12}. The mass
models predict rather small statistical uncertainties, but the
systematic uncertainties are at least 3 MeV. In the right panel, the
models are fit to a more limited data set in the Sn region, and while
the statistical uncertainties are nearly unchanged the systematic
uncertainties appear smaller. Of course, two models are insufficient
to make any conclusive statements about the magnitude of systematic
uncertainties, but this plot illustrates the basic point.

Constraints on the nuclear symmetry energy will be improved by
understanding this tension between completeness and accuracy. It might
be possible to obtain models which more accurately describe the
symmetry energy by fitting to only carefully selected nuclear data and
avoiding complications of attempting to describe nuclei which are
currently not well-described by Hartree-Fock-Bogoliubov. In more
detail, the best constraints on the symmetry energy will come from an
fitting several models to a data set which is specifically optimized
to generate the smallest combination of (i) the uncertainty in the fit
and (ii) the systematic uncertainty implied by the variation between
models. This optimization of this data set should include varying the
individual weights, $w_i$, in the $\chi^2$ function to ensure the
highest possible accuracy for the symmetry energy. In addition to
nuclear masses and charge radii described above, giant resonance data
will likely be helpful in determining the symmetry energy. When the
systematic uncertainties of neutron star mass and radius observations
are sufficiently small, they will also provide a powerful constraint.

\section{Discussion}

The analysis of neutron star data has a low computational requirement;
a single solution of the TOV equations is much faster than the
accurate determination of several nuclear masses and charge radii from
an energy density functional. This low computational requirement has
enabled the use of Bayesian analysis to attempt to understand
currently available neutron star data. A similar statistical analysis
for the matching of energy density functionals to low-energy nuclear
data is not yet possible. Nevertheless, the analogy permits
recommendations on future directions for fits to energy density
functionals.

\begin{itemize}
\item Likelihood functions should be published and made widely
  accessible.\footnote{Thanks to David W. Hogg for arguing
    that likelihoods ought to be published in his research blog.} If
  the likelihood function is sufficiently accurate for predictions,
  then publishing the best fit and the covariance matrix is
  sufficient. Otherwise, the likelihood function can either be
  tabulated or represented by a Markov chain of several points. In the
  neutron star case, this latter procedure was used in
  Ref.~\cite{Steiner14}; a list of points sampled from the likelihood
  function was provided for each model. This list can be directly used
  by other authors to provide predictions based on the same
  astronomical data. Also, the differences between the likelihood
  functions between models can be used to obtain an estimate on the
  systematic uncertainty as described above.
\item The classical $\chi^2$ procedure can be applied in cases where
  it is inappropriate, clouding our ability to properly interpret the
  data. Important methods to alleviate this issue include examining
  the residuals from the best fit model~\cite{Dobaczewski14}, and
  systematically refitting with part of the data set removed (as e.g.
  done in the neutron star context in Ref.~\cite{Guillot13}). In
  addition, one critical issue is the potential for correlations
  between our systematic uncertainties. More work needs to be done to
  understand how these correlations might potentially affect results.
  These correlations can be explicitly modeled using the method 
  described in Ref.~\cite{Stump01}. 
\item A method for obtaining modern constraints on the nuclear
  symmetry energy is proposed in the previous section. Generalizing
  this method, the ability of a model to make predictions should drive
  the data set which is included in $\chi^2$ fits and the associated
  weights that are used. This can only be effectively done when the
  systematic uncertainties of the model are assessed, either by a
  reasonable physical argument or a study of the variation among
  several models. The exception to this is when a more complete
  description of the data is likely to provide physical insight. The
  ability to make more accurate predictions will be useful, in
  particular, to those communities (like neutron star astronomers)
  who need nuclear data for their models. 
\end{itemize}

\section*{Acknowledgements}

The author would like to thank S. Liuti for pointing out
Ref.~\cite{Stump01}, G. Bertsch and S. Reddy for several
helpful discussions, and the DOE for support under grant
DE-FG02-00ER41132. Experimental values for the nuclear masses,
the theoretical mass formulas, and a solver for the stellar structure
equations were obtained from O2scl~\cite{o2scl}.

\section*{References}
\bibliographystyle{iopart-num}
\bibliography{paper}

\providecommand{\newblock}{}
\begin{thebibliography}{10}
\expandafter\ifx\csname url\endcsname\relax
  \def\url#1{{\tt #1}}\fi
\expandafter\ifx\csname urlprefix\endcsname\relax\def\urlprefix{URL }\fi
\providecommand{\eprint}[2][]{\url{#2}}

\bibitem{Lattimer01}
Lattimer J~M and Prakash M 2001 {\em Astrophys. J.\/} {\bf 550} 426
  \urlprefix\url{http://dx.doi.org/10.1086/319702}

\bibitem{Ozel09}
{{\"O}zel} F and {Psaltis} D 2009 {\em Phys. Rev. D\/} {\bf 80} 103003
  \urlprefix\url{http://dx.doi.org/10.1103/PhysRevD.80.103003}

\bibitem{Steiner10}
Steiner A~W, Lattimer J~M and Brown E~F 2010 {\em Astrophys. J.\/} {\bf 722} 33
  \urlprefix\url{http://dx.doi.org/10.1088/0004-637X/722/1/33}

\bibitem{Kass95}
Kass R~E and Raftery A~E 1995 {\em J. Am. Stat. Assoc.\/} {\bf 90} 773
  \urlprefix\url{http://dx.doi.org/10.1080/01621459.1995.10476572}

\bibitem{Lunney03}
Lunney D, Pearson J~M and Thibault C 2003 {\em Rev. Mod. Phys.\/} {\bf 75} 1021
  \urlprefix\url{http://dx.doi.org/10.1103/RevModPhys.75.1021}

\bibitem{Toivanen08}
Toivanen J, Dobaczewski J, Kortelainen M and Mizuyama K 2008 {\em Phys. Rev.
  C\/} {\bf 78}(3) 034306
  \urlprefix\url{http://dx.doi.org/10.1103/PhysRevC.78.034306}

\bibitem{Skyrme59}
Skyrme T~H~R 1959 {\em Nucl. Phys.\/} {\bf 9} 615
  \urlprefix\url{http://dx.doi.org/10.1016/0029-5582(58)90345-6}

\bibitem{Lalazissis97}
Lalazissis G~A, K\"onig J and Ring P 1997 {\em Phys. Rev. C\/} {\bf 55}(1)
  540--543 \urlprefix\url{http://link.aps.org/doi/10.1103/PhysRevC.55.540}

\bibitem{vWeizsacker35}
v~Weizs\"{a}cker C~F 1935 {\em Z. fur Physik\/} {\bf 96} 431
  \urlprefix\url{http://dx.doi.org/10.1007/BF01337700}

\bibitem{Bethe36}
Bethe H~A and Bacher R~F 1936 {\em Rev. Mod. Phys.\/} {\bf 8}(2) 82--229
  \urlprefix\url{http://dx.doi.org/10.1103/RevModPhys.8.82}

\bibitem{Duflo95}
Duflo J and Zuker A~P 1995 {\em Phys. Rev. C\/} {\bf 52} 23

\bibitem{Dieperink09}
Dieperink A~E~L and Van~Isacker P 2009 {\em Eur. Phys. J. A\/} {\bf 42} 269
  \urlprefix\url{http://dx.doi.org/10.1140/epja/i2009-10869-3}

\bibitem{Piekarewicz12}
Piekarewicz J, Agrawal B~K, Col\`o G, Nazarewicz W, Paar N, Reinhard P~G,
  Roca-Maza X and Vretenar D 2012 {\em Phys. Rev. C\/} {\bf 85}(4) 041302
  \urlprefix\url{http://link.aps.org/doi/10.1103/PhysRevC.85.041302}

\bibitem{Audi86}
Audi G, Davies W~G and Lee-Whiting G~E 1986 {\em Nucl. Instrum. Methods Phys.
  Res. A\/} {\bf 249} 443
  \urlprefix\url{http://dx.doi.org/10.1016/0168-9002(86)90700-X}

\bibitem{Audi03}
Audi G, Wapstra A~H and Thibault C 2003 {\em Nucl. Phys. A\/} {\bf 729} 337
  \urlprefix\url{http://dx.doi.org/10.1016/j.nuclphysa.2003.11.003}

\bibitem{Kondev14}
Kondev F 2014 priv. comm.

\bibitem{Michael94}
Michael C 1994 {\em Phys. Rev. D\/} {\bf 49} 2616--2619
  \urlprefix\url{http://dx.doi.org/10.1103/PhysRevD.49.2616}

\bibitem{Schunck14a}
Schunck N, McDonnell J~D, Higdon D, Sarich J and Wild S 2014 {\em
  arXiv:1406.4374\/} \urlprefix\url{http://arxiv.org/abs/1406.4374}

\bibitem{RocaMaza14}
Roca-Maza X, Paar N and Col\`{o} G 2014 {\em arXiv:1406.1885\/}
  \urlprefix\url{http://arxiv.org/abs/1406.1885}

\bibitem{Wild14}
Wild S~M, Sarich J and Schunck N 2014 {\em arXiv:1406.5464\/}
  \urlprefix\url{http://arxiv.org/abs/1406.5464}

\bibitem{Dobaczewski14}
Dobaczewski J, Nazarewicz W and Reinhard P~G 2014 {\em J. Phys. G.\/} {\bf 41}
  074001 \urlprefix\url{http://dx.doi.org/10.1088/0954-3899/41/7/074001}

\bibitem{Stump01}
Stump D, Pumplin J, Brock R, Casey D, Huston J, Kalk J, Lai H~L and Tung W~K
  2001 {\em Phys. Rev. D\/} {\bf 65} 014012
  \urlprefix\url{http://dx.doi.org/10.1103/PhysRevD.65.014012}

\bibitem{Coester70}
Coester F, Cohen S, Day B and Vincent C~M 1970 {\em Phys. Rev. C\/} {\bf 1}
  769--776 \urlprefix\url{http://dx.doi.org/10.1103/PhysRevC.1.769}

\bibitem{Lattimer14a}
Lattimer J~M and Steiner A~W 2014 {\em Eur. Phys. J. A\/} {\bf 50} 40
  \urlprefix\url{http://dx.doi.org/10.1140/epja/i2014-14040-y}

\bibitem{Alford13}
Alford M~G, Han S and Prakash M 2013 {\em Phys. Rev. D\/} {\bf 88} 083013
  \urlprefix\url{http://dx.doi.org/10.1103/PhysRevD.88.083013}

\bibitem{Gelman12}
Gelman A and Shalizi C~R 2012 {\em Brit. Jour. Math. Stat. Psych.\/} {\bf 66} 8
  \urlprefix\url{http://dx.doi.org/10.1111/j.2044-8317.2011.02037.x}

\bibitem{Steiner14}
Steiner A~W, Gandolfi S, Fattoyev F~J and Newton W~G 2014 {\em
  arXiv:1403.7546\/} \urlprefix\url{http://arxiv.org/abs/1403.7546}

\bibitem{Hebeler13}
Hebeler K, Lattimer J~M, Pethick C~J and Schwenk A 2013 {\em Astrophys. J.\/}
  {\bf 773} 11 \urlprefix\url{http://dx.doi.org/10.1088/0004-637X/773/1/11}

\bibitem{Steiner13}
Steiner A~W, Lattimer J~M and Brown E~F 2013 {\em Astrophys. J. Lett.\/} {\bf
  765} 5 \urlprefix\url{http://dx.doi.org/10.1088/2041-8205/765/1/L5}

\bibitem{Steiner12}
{Steiner} A~W and Gandolfi S 2012 {\em Phys. Rev. Lett.\/} {\bf 108} 081102
  \urlprefix\url{http://dx.doi.org/10.1103/PhysRevLett.108.081102}

\bibitem{Lattimer14b}
Lattimer J~M and Steiner A~W 2014 {\em Astrophys. J.\/} {\bf 784} 123
  \urlprefix\url{http://dx.doi.org/10.1088/0004-637X/784/2/123}

\bibitem{Heinke14}
Heinke C~O, Cohn H~N, Lugger P~M, Webb N~A, Ho W~C~G, Anderson J, Campana S,
  Bogdanov S, Haggard D, Cool A~M and Grindlay J~E 2014 {\em Mon. Not. Roy.
  Astron. Soc.\/} {\bf 444} 443
  \urlprefix\url{http://dx.doi.org/10.1093/mnras/stu1449}

\bibitem{Kortelainen12}
Kortelainen M, McDonnell J, Nazarewicz W, Reinhard P~G, Sarich J, Schunck N,
  Stoitsov M~V and Wild S~M 2012 {\em Phys. Rev. C\/} {\bf 85}(2) 024304
  \urlprefix\url{http://dx.doi.org/10.1103/PhysRevC.85.024304}

\bibitem{Farine78}
Farine M, Pearson J~M and Rouben B 1978 {\em Nucl. Phys. A\/} {\bf 304} 317
  \urlprefix\url{10.1016/0375-9474(78)90241-5}

\bibitem{Horowitz01}
Horowitz C~J and Piekarewicz J 2001 {\em Phys. Rev. Lett.\/} {\bf 86}
  5647--5650 \urlprefix\url{http://dx.doi.org/10.1103/PhysRevLett.86.5647}

\bibitem{Abrahamyan12}
{Abrahamyan S et al (PREX collaboration)} 2012 {\em Phys. Rev. Lett.\/} {\bf
  108} 112502 \urlprefix\url{http://dx.doi.org/10.1103/PhysRevLett.108.112502}

\bibitem{Tarbert14}
{Tarbert C M et al (Crystal Ball at MAMI and A2 Collaboration)} 2014 {\em Phys.
  Rev. Lett.\/} {\bf 112} 242502
  \urlprefix\url{http://dx.doi.org/10.1103/PhysRevLett.112.242502}

\bibitem{Hebeler10}
Hebeler K, Lattimer J~M, Pethick C~J and Schwenk A 2010 {\em Phys. Rev.
  Lett.\/} {\bf 105}(16) 161102
  \urlprefix\url{http://dx.doi.org/10.1103/PhysRevLett.105.161102}

\bibitem{Furnstahl02}
Furnstahl R~J 2002 {\em Nucl. Phys. A\/} {\bf 706} 85
  \urlprefix\url{http://dx.doi.org/10.1016/S0375-9474(02)00867-9}

\bibitem{Stone03}
Stone J~R, Miller J~C, Koncewicz R, Stevenson P~D and Strayer M~R 2003 {\em
  Phys. Rev. C\/} {\bf 68} 034324
  \urlprefix\url{http://dx.doi.org/10.1103/PhysRevC.68.034324}

\bibitem{Steiner05}
Steiner A~W, Prakash M, Lattimer J~M and Ellis P~J 2005 {\em Phys. Rep.\/} {\bf
  411} 325 \urlprefix\url{http://dx.doi.org/10.1016/j.physrep.2005.02.004}

\bibitem{Epelbaum09}
Epelbaum E, Hammer H~W and Mei\ss{}ner U~G 2009 {\em Rev. Mod. Phys.\/} {\bf
  81} 1773--1825 \urlprefix\url{http://dx.doi.org/10.1103/RevModPhys.81.1773}

\bibitem{Hagen12}
Hagen G, Hjorth-Jensen M, Jansen G~R, Machleidt R and Papenbrock T 2012 {\em
  Phys. Rev. Lett.\/} {\bf 109} 032502
  \urlprefix\url{http://dx.doi.org/10.1103/PhysRevLett.109.032502}

\bibitem{Wienholtz13}
Wienholtz F, Beck D, Blaum K, Borgmann C, Breitenfeldt M {\em et~al.\/} 2013
  {\em Nature\/} {\bf 498} 346--349
  \urlprefix\url{http://dx.doi.org/10.1038/nature12226}

\bibitem{Hagen13}
Hagen G, Papenbrock T, Hjorth-Jensen M and Dean D~J 2014 {\em Rep. Prog.
  Phys.\/} {\bf 77} 096302
  \urlprefix\url{http://dx.doi.org/10.1088/0034-4885/77/9/096302}

\bibitem{Steiner12b}
Steiner A~W 2012 {\em Phys. Rev. C\/} {\bf 85} 055804
  \urlprefix\url{http://dx.doi.org/10.1103/PhysRevC.85.055804}

\bibitem{Audi12}
Audi G, Wang M, Wapstra A~H, Kondev F~G, MacCormick M, Xu X and Pfeiffer B 2012
  {\em Chin. Phys. C\/} {\bf 36} 1287
  \urlprefix\url{http://dx.doi.org/10.1088/1674-1137/36/12/002}

\bibitem{Wang12}
Wang M, Audi G, Wapstra A~H, Kondev F~G, MacCormick M, Xu X and Pfeiffer B 2012
  {\em Chin. Phys. C\/} {\bf 36} 1603
  \urlprefix\url{http://dx.doi.org/10.1088/1674-1137/36/12/003}

\bibitem{Guillot13}
Guillot S, Servillat M, Webb N~A and Rutledge R~E 2013 {\em Astrophys. J.\/}
  {\bf 772} 7 \urlprefix\url{http://dx.doi.org/10.1088/0004-637X/772/1/7}

\bibitem{o2scl}
Steiner A~W 2014 O2scl, object-oriented scientific computing library record
  ascl:1408.019
  \urlprefix\url{http://adsabs.harvard.edu/abs/2014ascl.soft08019S}

\end{thebibliography}

\end{document}